\documentclass[aps,preprint,nofootinbib,preprintnumbers,superscriptaddress,eqsecnum]{revtex4-1}
\usepackage{graphics,epsfig,color,subfigure}
\usepackage{amssymb}
\usepackage{physics}
\usepackage{subfig}
\usepackage{dirtytalk}

\usepackage{subfig}

\newcommand{\Ket}[1]{{\bf \lbrack #1\rangle} }

\newcommand{\SWAP}{{\bf SWAP}}
\newcommand{\Xo}{{\bf X}}

\newcommand{\Zo}{{\bf Z}}

\begin{document}

\title{Fast Scrambling in Classically Simulable Quantum Circuits}
\author{Mike Blake}
\affiliation{School of Mathematics, 
University of Bristol, Fry Building, \\Woodland Road, Bristol BS8 1UG, UK }
\author{Noah Linden}
\affiliation{School of Mathematics, 
University of Bristol, Fry Building, \\Woodland Road, Bristol BS8 1UG, UK }
\author{Anthony P. Thompson}
\affiliation{School of Mathematics, 
University of Bristol, Fry Building, \\Woodland Road, Bristol BS8 1UG, UK }
\affiliation{Quantum Engineering Center for Doctoral Training, University of Bristol, \\ Tyndall Avenue, Bristol BS8 1FD, UK}

 \begin{abstract}
 
 We study operator scrambling in quantum circuits built from `super-Clifford' gates. For such circuits it was established in~\cite{Blake_2020} that the time evolution of operator entanglement for a large class of many-body operators can be efficiently simulated on a classical computer, including for operators with volume-law entanglement. Here we extend the scope of this formalism in two key ways. Firstly we provide evidence that these classically simulable circuits include examples of fast scramblers, by constructing a circuit for which operator entanglement is numerically found to saturate in a time $t_* \sim \mathrm{ln}(N)$ (with $N$ the number of qubits). Secondly we demonstrate that, in addition to operator entanglement, certain out-of-time ordered correlation functions (OTOCs) can be classically simulated within the same formalism. As a consequence such OTOCs can be computed numerically in super-Clifford circuits with thousands of qubits, and we study several explicit examples in the aforementioned fast scrambling circuits.
    \end{abstract}

\maketitle
\tableofcontents

\section{Introduction} A fundamental feature of chaotic many-body quantum systems is the scrambling of quantum information. One particular probe of such chaotic dynamics is operator scrambling - how an initially simple operator $W(0)$ becomes increasingly complex under Heisenberg time evolution $W(t) = U^{\dagger}(t) W(0) U(t)$, with $U(t)$ the time evolution operator.  Several diagnostics have been introduced to quantify the phenomenon of operator scrambling, both in Hamiltonian systems and in quantum circuit models. Perhaps the most well known are the decay of out-of-time ordered correlation functions (OTOCs)  \cite{Nahum_2017, Nahum_2018, Khemani_2018, vonKeyserlingk_2018, vonKeyserlingk_20182, Roberts_2015, RobertsStanford2018, Shenker_2014, Maldacena_2016, Kitaev_2018, Blake_2018, Hosur_2016} and the generation of operator entanglement (see, for example, \cite{Zanardi:2001zza, Prosen_2007, Pizorn:2009gup, Prosen_2008, Jonay:2018yei, Mezei:2019zyt, Zhou:2016wpr, Nie:2018dfe, Pal:2018nvj, PhysRevB.100.214301, Alba:2019okd}). Although less familiar than state entanglement, the operator entanglement of an operator $W(t)$ acting on a Hilbert space $\mathcal{H}$ can formally be defined by identifying $W(t)$ with a suitably normalized state in a larger Hilbert space. Given a (spatial) bipartition $\mathcal{H} = \mathcal{H}_A \otimes \mathcal{H}_{\bar{A}}$ the operator entanglement then simply refers to the entropy of the corresponding state in operator space across the bipartition. 

Our current understanding of operator scrambling in generic systems is limited by the computational challenge of tracking the Heisenberg time evolution of operators in a many-body quantum system. In particular, in a system of $N$ qubits expanding an operator $W(t)$ in a basis of Pauli strings gives rise to exponentially many $(4^{N})$ time dependent amplitudes to track. Surprisingly then \cite{Blake_2020} introduced examples of quantum circuits, so-called `super Clifford' circuits, for which the generation of long-range (volume-law) operator entanglement can be efficiently simulated on a classical computer, in a time polynomial in $N$. This is one of the few known numerical techniques for studying operator scrambling in quantum circuits with thousands of qubits, see \cite{PhysRevB.100.214301, Alba:2019okd, Gopalakrishnan:2018rfu, Gopalakrishnan:2018wqv} for an alternative method based on cellular automata. 

The techniques introduced in \cite{Blake_2020}, which we review in Section~\ref{supercliff}, worked by identifying unitary gates whose action on a subspace $\mathcal{S}$ of operators was analogous to the action of Clifford gates on quantum states.  For such `super-Clifford' gates, the operator entanglement of operators $W(t) \in \mathcal{S}$ could be extracted by tracking a linear (in $N$) number of `super-stabilizers', which provide a highly efficient representation of the time evolved operator.  Such circuits provide a new route to exploring the scrambling of operators in large many-body systems, which for the aforementioned reasons remains challenging outside of certain analytically soluble models. In this paper we continue the exploration of scrambling in super-Clifford circuits, going beyond the results of~\cite{Blake_2020} in two key ways. 

Firstly a central question in the context of scrambling, that was not discussed in detail in \cite{Blake_2020}, is to understand how quickly operator entanglement can be generated by super-Clifford circuits.  A key motivation for this question is provided by the fast scrambling conjecture \cite{Sekino_2008} which hypothesised a lower bound on the `scrambling time' $t_*$ of many-body quantum systems with few-body interactions. Systems which scramble information as quickly as possible, so-called `fast scramblers', have scrambling time \cite{Hayden_2007, Sekino_2008}
\begin{equation}
t_* \sim  \ln(N)
\label{fastscramblerintro}
\end{equation}
%
%
In Section~\ref{sec:entanglement} we study the scrambling time in a particular `parallel-processing' super-Clifford circuit, whose structure is motivated by random unitary models that inspired the fast-scrambling conjecture. We extract the scrambling time $t_*$ as the time taken for the operator entanglement of an initial product operator to saturate on a macroscopic subregion. We find strong numerical evidence that the scrambling time scales precisely as \eqref{fastscramblerintro} for large $N$, i.e. that the system is a fast scrambler with regards to the generation of operator entanglement.

Secondly, in Section~\ref{sec:OTOC} we significantly expand the types of probes of scrambling that can be efficiently computed in super-Clifford circuits.  In particular, we demonstrate that certain out-of-time ordered correlation functions (OTOCs) involving $W(t) \in \mathcal{S}$ can be computed using the `super-stabiliser formalism'. As a result, such OTOCs can be computed in polynomial time, and hence in super-Clifford circuits can be numerically studied in systems of thousands of qubits. We also provide explicit numerical results for several OTOCs in the aforementioned `parallel-processing' super-Clifford circuit, and demonstrate that they exhibit the expected features of scrambling of $W(t)$ in $\mathcal{S}$. 

\section{Super-Clifford circuits}
\label{supercliff} 

Super-Clifford circuits, introduced in \cite{Blake_2020}, are a class of quantum circuits for which the Heisenberg time evolution of a (sub-class) of operators $W(t)$ can be efficiently simulated. 
In particular, in a Hilbert space of $N$ qubits, we consider the subspace of operators spanned by strings consisting of an $X$ or a $Y$ at each site (with $X$, $Y$ the usual Pauli operators). We will refer to this $2^N$ dimensional subspace of operators as $\mathcal{S}$. The `super-Clifford' circuits introduced in \cite{Blake_2020} have two essential properties. Firstly, under super-Clifford dynamics, the time evolution of operators within the $2^{N}$ dimensional subspace $\mathcal{S}$ is closed. Furthermore, the action (by conjugation) of super-Clifford gates on operators in $\mathcal{S}$ is equivalent to the action of certain Clifford gates on states.  As a consequence, the dynamics of operators $W(t) \in \mathcal{S}$ can be studied by adapting techniques used to simulate Clifford circuits to operator space. 

In order to demonstrate the connection to Clifford dynamics it is convenient to introduce a state like notation to represent operators in $\mathcal{S}$. In particular we denote $X = \Ket{0}$ and $Y = \Ket{1}$ such that the operator $X_1X_2Y_3$ in a system of 3 qubits is represented by the notation $\Ket{001}$. We note that this notation the operator entanglement of an operator $W \in \mathcal{S}$ across a spatial bipartition of $\mathcal{H}$ is simply the entanglement entropy of the state $\Ket{W}$, such that for an operator $W \in \mathcal{S}$ the maximum amount of operator entanglement across a bipartition $A, \bar{A}$ with $d_{A} < d_{\bar{A}}$ is $S_{A} = \mathrm{log}_2 d_A = N_A$. Here $d_A = 2^{N_A}$ is the Hilbert space dimension of $A$.

A simple example of a super-Clifford gate is then provided by the familiar $T$ gate. This acts on operators in ${\mathcal S}$ by conjugation as
\begin{equation}
    T^{\dagger} X T = \frac{X -Y}{\sqrt{2}}, \hspace{1cm} T^{\dagger}Y T = \frac{X + Y}{\sqrt{2}}. 
\end{equation}
In state-like notation can be written as the action
\begin{equation}
\Ket{0} \to \frac{\Ket{0} -\Ket{1}}{\sqrt{2}}, \hspace{1cm} \Ket{1} \to \frac{\Ket{0} + \Ket{1}}{\sqrt{2}},
\end{equation}
which is equivalent to that of a super-operator Hadamard $\mathbf{H}$ followed by the super-operator $\mathbf{Z}$ (here by super-operators we mean linear maps acting on $\mathcal{S}$). A second example is the familiar $\mathrm{SWAP}$ gate, interchanging the states of two qubits. The action of $\mathrm{SWAP}$ on operators is entirely analogous, for instance we have:
\begin{equation}
    \mathrm{SWAP^{\dagger}}{X_1 Y_2}\mathrm{SWAP} = Y_1 X_2
\end{equation}
which we can write in state-like notation as
\begin{equation}
    \mathbf{SWAP}\Ket{01} = \Ket{10}
\end{equation}
A more non-trivial example of a super-Clifford gate, that is crucial to generating operator entanglement, was identified in ref. \cite{Blake_2020} as
\begin{equation}
    \mathrm{C3} = \mathrm{CX}_{21}\mathrm{CX}_{31}\mathrm{CZ}_{12}T_{1}^6 T_2^6
\end{equation}
whose action on the subspace $\mathcal{S}$ is equivalent to a product of control gates  $\mathbf{C_3} = \mathbf{CY}_{12}\mathbf{CY}_{13}$.

In summary, the time evolution of operators in $\mathcal{S}$ in a circuit built from the gates $T, \mathrm{SWAP}, C3$ is equivalent to evolving the corresponding state $\Ket{W}$ under Clifford dynamics generated by $\mathbf{Z} \mathbf{H}, \mathbf{SWAP}, \mathbf{C_3}$.  It was demonstrated in~\cite{Blake_2020} that super-Clifford circuits built from this gate-set can lead to volume law operator entanglement, i.e. operators $W(t)$ with $S_A \approx N_A$, starting from an initial (unentangled) product operator $W(0) = X_1 \dots X_N $. Furthermore, the generation of operator entanglement could be classically simulated in polynomial time by adapting the stabiliser formalism to the super-Clifford setting. We will also demonstrate in Section~\ref{sec:OTOC}  that the super-stabiliser formalism can be used to compute certain OTOCs involving $W(t)$ in polynomial time. As such, we now review how the stabiliser formalism~\cite{Aaronson_2004} generalises to the setting of super-Clifford circuits, following~\cite{Blake_2020}.

In the context of our super-Clifford circuits, the familiar notion of a stabiliser state is replaced by a `stabiliser operator'. Specifically, for an $N$ qubit system, we define a stabiliser operator as an operator in $\mathcal{S}$ which is fixed under the action of $N$ independent super-Pauli strings - the so-called `super-stabilizers'. In our state-like notation we have $N$ super-stabilisers $\mathbf{O}_{\alpha}$ such that $\mathbf{O}_{\alpha} \Ket{W} = \Ket{W}$. The super-stabilisers of $\Ket{W}$ form a $2^{N}$ dimensional group under multiplication, and by independent we mean that they form a choice of generating set for the super-stabiliser group, and none can be generated from the others. A stabiliser operator is uniquely defined by its super-stabiliser group, and vice versa. 

A simple example of a super-stabiliser operator is a product of $X$s and $Y$s, i.e. a computational basis state in terms of $\Ket{0}$s and $\Ket{1}$s. For such an operator the super-stabilisers can be chosen to be 
\begin{equation}
\mathbf{\hat{O}}_{\alpha} = (-1)^{s_{\alpha}}  \mathbf{Z}_{\alpha},
\label{compbasisstab}
\end{equation}
where $\alpha = 1, \dots, N$ and $s_{\alpha} = 0 $ if there is an $X$ at site ${\alpha}$ and $1$ otherwise. In general for an Hermitian stabiliser operator the $N$ super-stabilisers can be decomposed as \footnote{We note in \cite{Blake_2020} we used the notation $\mathbf{v}_{1x}$ instead of $v_{1x}$.}
\begin{equation}
    \mathbf{O}_{\alpha} = (-1)^{s_{\alpha}} \mathbf{X_1}^{v_{1 x}} \mathbf{Z_1}^{v_{1 z}} ... \mathbf{X_N}^{v_{N x}} \mathbf{Z_N}^{v_{N z}},
    \label{stab_form}
\end{equation}
with $\mathbf{X_i}, \mathbf{Z_i}$ super-Pauli operators (i.e. Pauli operators acting on the space $\mathcal{S}$). 

We now consider starting from an initial product operator $W(0)$, with initial super stabilisers~\eqref{compbasisstab}, and evolving with a super-Clifford circuit $U = U_{\tau}... U_{1}$ - i.e. where each $U_i$ has been selected from the gate set $\{T, \mathrm{SWAP}, C3\}$. Under such a circuit the operator $W(\tau)$ remains a stabiliser operator with super-stabilisers given by 
\begin{equation}
    \mathbf{O}_{\alpha}(\tau) = \mathbf{U}_1...\mathbf{U}_\tau \mathbf{\hat{O}}_\alpha \mathbf{U}_\tau^\dagger ... \mathbf{U}^{\dagger}_1, 
\end{equation}
with $\mathbf{U}_i$ the super-operator describing the action of the gate $U_i$ on $\mathcal{S}$.  Note the property that $\{T, \mathrm{SWAP}, C3\}$ act as Clifford gates on $\mathcal{S}$ is necessary for $W(\tau)$ to remain a stabiliser operator, since it implies that under conjugation each $\mathbf{U}_i$ maps a super-Pauli string to a single super-Pauli string. As such under a super-Clifford circuit the super-stabilisers remain of the form~\eqref{stab_form} and can be efficiently tracked simply by updating the $N$ binary vectors
\begin{equation}
    \mathbf{v}_{\alpha} = (v_{1x}, v_{1z},..., v_{N x}, v_{N z})
    \label{vector}
\end{equation}
and the overall sign $s_{\alpha}$ of each super-stabiliser under the action of each gate. 

For our super-Clifford circuit, we then need to understand how the gates $T$, $\mathrm{SWAP}$ and $C3$ act on the super-stabiliser vectors \eqref{vector}. The gate $T$ acting on the $i$th qubit acts on the super-stabilisers as 
\begin{equation}{\bf X_i} \to {\bf Z_i}; \;\;\; {\bf Z_i} \to -{\bf X_i}
\label{Tupdate}
\end{equation}
which can be tracked by exchanging $v_{ix}$ and $v_{iz}$ in \eqref{vector} and updating the overall sign of the stabilisers through
\begin{equation}
s_{\alpha} \to s_{\alpha} + f(v_{i x}, v_{i z}). 
\end{equation}
where $f(v_{ix}, v_{iz}) = 1$ if $v_{ix} = 0, v_{iz} = 1$ and $f(v_{ix}, v_{iz})= 0$ else. The operator ${\bf C3}$ acts by
\begin{eqnarray} 
 \Xo_1 &\to&  - \Xo_1 \Xo_2 \Zo_2 \Xo_3 \Zo_3, \;\; \Xo_2 \to \Zo_1\Xo_2, \;\; \Xo_3 \to \Zo_1 \Xo_3, \nonumber \\
 \Zo_1 &\to&  \Zo_1, \;\;   \Zo_2 \to \Zo_1 \Zo_2,  \;\; \Zo_3\to  \Zo_1\Zo_3, 
 \label{C3update}
\end{eqnarray}
which updates the binary vectors ${\bf v}_{\alpha}$ according to %
\begin{eqnarray}
\label{controln}
{v}_{1x} &\to & {v}_{1x}, \;\; {v}_{1z} \to {v}_{1z} + {v}_{2x}+{v}_{2z} + {v}_{3x}+{ v}_{3z} ,\nonumber\\
{ v}_{2x} &\to & {v}_{1x} + {v}_{2x} ,\;\; {v}_{2z} \to {v}_{1x}+ { v}_{2z},\nonumber\\
{ v}_{3x} &\to & { v}_{1x} + { v}_{3x} ,\;\; { v}_{3z} \to {v}_{1x} + {v}_{3z}. 
\end{eqnarray}
as well as the overall signs of the stabilisers by
\begin{equation}
s_{\alpha} \to s_{\alpha} + v_{1x} + \sum_{i=2}^{3} g(v_{1x}, v_{ix})
\end{equation}
where $g(v_{1x}, v_{ix}) = 1$ if $v_{1x} = 1, v_{ix} = 1$ and $g(v_{1x}, v_{ix})= 0$ else. 
Finally $\SWAP$ acts by exchanging the components of $\bf{v}_{\alpha}$ for the two qubits on which it acts. 

The operator entanglement $S_A(\tau)$ can be directly extracted from the updated super-stabiliser vectors~\eqref{vector} using the techniques of \cite{Nahum_2017}. Specifically, one considers the matrix formed by combining the $N$ super-stabilisers into a $2 N$ by $N$ matrix ${\bf V} = ({\bf v_1}^{T}, \dots, {\bf v_N}^{T})$. In terms of this matrix of super-stabilisers then the operator entanglement entropy (with base $2$ logarithm) of a subregion $A$ consisting of the first $N_A$ qubits is given by $S_A(\tau) = I_A - N_A$, where $I_A$ is the rank (in arithmetic modulo 2) of the submatrix formed by keeping the first $2 N_A$ rows of ${\bf V}$.

It is interesting to note that the operator entanglement entropy does not depend on the overall signs of the stabilisers $s_{\alpha}$ - this is in contrast to OTOCs, which we will see in Section~\ref{sec:OTOC} can also be computed in the stabiliser formalism but are sensitive to $s_{\alpha}$.  An immediate consequence of this observation is that under super-Clifford dynamics  the operator entanglement $S_{A}(\tau)$ will be the same for any choice of computational basis state as the initial operator $W(0)$ (since the super-stabilisers for such operators~\eqref{compbasisstab} differ only by possible signs).

\section{Fast Scrambling of Operators}
\label{sec:entanglement}
 
The previous work \cite{Blake_2020} performed initial studies of operator entanglement in super-Clifford circuits, and in particular demonstrated that super-Clifford circuits were capable of generating a large amount of operator entanglement starting from an initial product operator $W(0) = X_1 \dots X_N$. We note that this implies that the physics of such circuits is qualitatively distinct from conventional Clifford circuits. In particular whilst Clifford circuits are capable on generating state entanglement from an initial product state, they are not capable of generating operator entanglement starting from a single Pauli string. 

A central question in the context of scrambling, that was not discussed in detail in \cite{Blake_2020}, is to understand how quickly operator entanglement can be generated by super-Clifford circuits. A key motivation for this question is provided by the fast scrambling conjecture \cite{Sekino_2008}, which hypothesised a lower bound on the `scrambling time' $t_*$ of many-body quantum systems - that is the time taken for the system to scramble initially local quantum information across many degrees of freedom. Specifically, in a system of $N$ qubits, \cite{Sekino_2008} introduced the concept of a `fast scrambler', i.e. a system which scrambles in a time 
\begin{equation}
t_* \sim  \ln(N)
\label{fastscrambler}
\end{equation}
in the limit $N \to \infty$. Further~\eqref{fastscrambler} was conjectured to represent a lower on bound on the scrambling time of generic many-body quantum systems with few-body interactions~\cite{Sekino_2008}. Examples of fast-scrambling systems include certain `parallel-processing' random unitary circuits, holographic quantum field theories and the SYK model \cite{Hayden_2007, Sekino_2008, Shenker_2014, Maldacena_2016, Kitaev_2018}. 

In this section we study the scrambling time of a particular `parallel-processing' super-Clifford circuit, whose structure is inspired by random unitary models  \cite{Hayden_2007, Sekino_2008, dankert2005efficientsimulationrandomquantum, Dankert_2009}  that motivated the fast scrambling conjecture. We define the scrambling time $t_*$ as the time taken for the operator entanglement of an initial product operator to saturate. We will provide compelling numerical evidence that at large $N$ the scrambling time scales as $t_* \sim \mathrm{ln}(N)$ - i.e. that this super-Clifford circuit is a fast scrambler with regards to the generation of operator entanglement.

\subsection{Scrambling time from operator entanglement}

To proceed further it is necessary to give a precise definition of how we extract a scrambling time from operator entanglement. To do this, we generalise a natural definition introduced in~\cite{Bentsen_2019} in the context of state entanglement. In this setting, the authors of~\cite{Bentsen_2019} considered the entanglement entropy of an initial product state evolved under chaotic quantum dynamics. The scrambling time was then defined as the time taken for the entanglement entropy of a macroscopic subregion $A$ to (almost) saturate. A particularly appealing aspect of this definition of $t_*$ is that in this setting one can give a rigorous proof of the fast scrambling conjecture in Hamiltonian systems with exponentially decaying two-point functions~\cite{Bentsen_2019}.

In our context we can then define a scrambling time associated to the generation of operator entanglement by a super-Clifford circuit, in analogy to~\cite{Bentsen_2019}. Specifically, we consider starting from an initial computational basis (product) operator $W(0) \in {\cal S}$ and evolve in time under a super-Clifford circuit. We take $A$ to be an arbitrary fixed fraction $m < 1/2$ of the qubits (such that $N_A = m N$) and consider the operator entanglement entropy $S_m(t)$ of the operator $W(t)$ on the subregion $A$. At late times, under a suitably chaotic circuit, the operator entanglement entropy will approach that of a maximally mixed operator in $A$, such that $S_m(t) = S^{\mathrm{sat}}_m \equiv m N$. We then define the scrambling time $t_*$ as the time taken for operator entanglement entropy $S_m(t)$ to be within a fixed ($N$-independent) amount of the saturation value, i.e. the smallest time $t$ such that:
\begin{equation}
S_m(t) \geq S^{\mathrm{sat}}_m - \epsilon
\label{scrambling_time}
\end{equation}
with $\epsilon > 0$ a fixed constant independent of $N$. This definition of scrambling time is entirely analogous to that used in~\cite{Bentsen_2019} for state entanglement\footnote{We note that it is also common to define a scrambling time in terms of OTOCs of local operators, e.g. as in ~\cite{PhysRevB.103.L121111, PhysRevE.99.052212}. We are not aware of a direct relationship between these alternative definitions and those based on entanglement entropy as used here or in \cite{Bentsen_2019}.}.

\subsection{A fast scrambling super-Clifford circuit}
\label{sec:fastscrambler}

We now wish to describe the particular super-Clifford circuit we will study in this paper, motivated by random circuit models which inspired the fast scrambling conjecture. It is instructive to first recall a family of random super-Clifford circuits studied in~\cite{Blake_2020}.  There, one considered evolving the operator $X_1X_2...X_N \equiv \Ket{00...0}$ by acting with a two step circuit. Firstly, one acts on a randomly drawn qubit with the $T$ gate. Secondly, one randomly draws a qubit $j \in {1,..., N-2}$ and acts with the gate $C3$ on qubits $j, j+1, j+2$, randomizing which of these will act as control. This whole operation counts as a single time step, and led to a circuit for which operator entanglement saturated in a time of order $\mathcal{O}(N^2)$.

In order to build circuits that scramble operators more efficiently we modify the circuit discussed in~\cite{Blake_2020} in two ways, inspired by `parallel processing' quantum circuit models discussed in~\cite{Dankert_2009, Hayden_2007, Sekino_2008} that motivated the original fast-scrambling conjecture. Firstly, we now act in each timestep with both ${\cal O}(N)$ $T$ and ${\cal O}(N)$ $C3$ gates (on distinct qubits). Secondly, we consider a model with all-to-all rather than nearest neighbour interactions - that is the three sites the multi-qubit gate $C3$ acts on are now chosen randomly (but distinctly). 

To be concrete, the simulations presented in this paper were performed for the following circuit, which is defined for system sizes $N$ which are divisible by $40$: 
\begin{enumerate}
    \item Start with $X_1X_2...X_N \equiv \Ket{00...0}$
    \item Randomly draw a set $\Gamma \subset \{1, ... , N\}$ with $|\Gamma| = N/10  $.
    \item Randomly draw $3 |\Gamma|/4 $ of the qubit labels in $\Gamma$ and act with $C3$ on these qubits (in randomly chosen groups of 3 qubits).
    \item Act with $T$ on the remaining qubits in $\Gamma$.
    \item Steps 2-4 count as a single time-step. Repeat, over many time-steps denoted by $t$. 
\end{enumerate}

The time evolution of stabilisers under this circuit can be tracked directly using the results in Section II, or by making use of existing software packages for simulating Clifford dynamics. For the explicit numerical results presented in this paper we have used the Clifford simulation software `stim' \cite{Gidney_2021} . We have checked that similar numerical results are obtained in other random circuits that apply $T$ gates and $C3$ gates in a qualitatively similar manner.

\subsection{Operator entanglement entropy}

\begin{figure}[h!]
\centerline{\includegraphics[scale= 0.50]{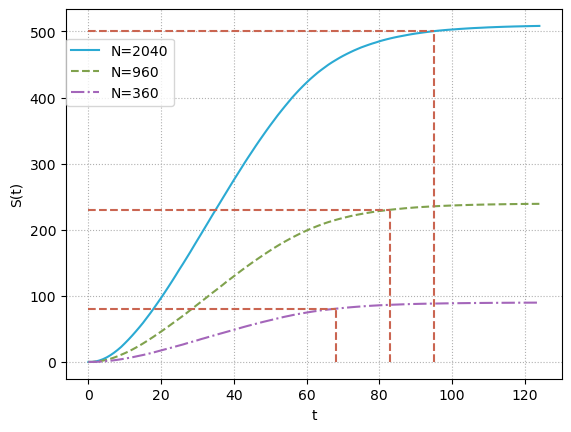}}
\caption{The operator entanglement entropy $S_{1/4}(t)$ (with base $2$ logarithm) of a quarter of the qubits for a parallel-processing super-Clifford circuit, averaged over $500$ iterations of the circuit. The different plots correspond to different numbers of qubits $N$. The dotted lines indicate the extraction of the scrambling time using the definition~\eqref{scrambling_time}.}
\label{entropyplot}
\end{figure}

Here we present numerical results for the operator entanglement entropy in the random circuit model introduced in Section~\ref{sec:fastscrambler}, with the goal of understanding the scaling of the scrambling time for large $N$. In order to smooth out circuit-to-circuit fluctutations the results displayed in the main text are for the entanglement entropy $S_m(t)$ averaged over many realisations of our random circuit. The scrambling time is then extracted from $S_m(t)$ using the definition~\eqref{scrambling_time}. We demonstrate in Appendix~\ref{app:scramblingtime} that similar results are obtained if one instead averages the scrambling times computed for individual realisations of the circuit.  

For illustrative purposes, numerical plots of the averaged operator entanglement entropy of a quarter of the system, $m=1/4$, are shown in Figure~\ref{entropyplot}. The dotted lines in Figure~\ref{entropyplot} show the scrambling time extracted by the definition~\eqref{scrambling_time}, which can be seen to be an increasing function of $N$. We can now extract the large-$N$ scaling of the scrambling time in two distinct ways. Firstly, we demonstrate numerically that for large $N$ the near-saturation behaviour of the averaged operator entanglement entropy is well described by a simple scaling form
\begin{figure*}[htp]
\centering
\includegraphics[width=.5\textwidth]{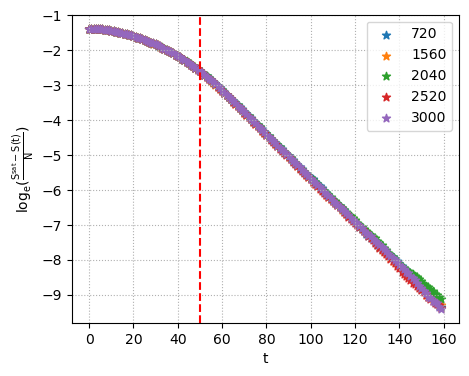}\hfill
\includegraphics[width=.5\textwidth]{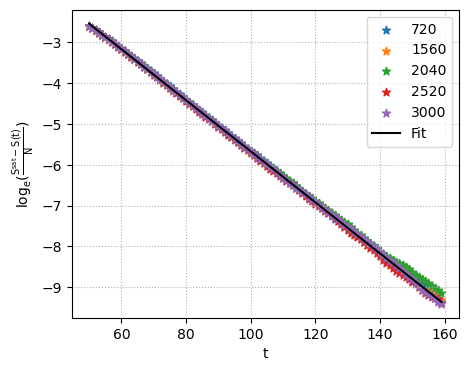}\hfill
\caption{(Left) $\mathrm{ln}(\Delta(S_{1/4}(t)))$ for a range of system sizes $N\geq 720$ averaged over $500$ iterations of a parallel-processing super-Clifford circuit. At an N independent timescale (shown by the red dotted line) the data is approximately described by a straight line with N independent parameters. (Right) The same data but only for times $t \geq 50$, including a straight line fit (black line) on the data obtained for $N=3000$.  The straight line with independent parameters observed here is consistent with the scaling form~\eqref{func_form} implying the system is fast scrambling.}
\label{StraightLine}
\end{figure*}
\begin{equation}
    S_{m}(t) = S^{\mathrm{sat}}_m - \alpha N e^{-\lambda t}
\label{func_form}
\end{equation}
where $S^{\mathrm{sat}}_m \equiv m N$ and $\alpha, \lambda$ are $N$ independent constants that in principle can depend on the fraction $m$ of the subsystem we are studying. It immediately follows from the functional form~\eqref{func_form} and the definition~\eqref{scrambling_time} that the scrambling time scales as~\eqref{fastscrambler}. In particular, letting $t_*$ denote the scrambling time according to~\eqref{scrambling_time} then the scaling form~\eqref{func_form} implies 
\begin{equation}
    \alpha Ne^{-\lambda t_{*}} = \epsilon 
    \implies t_* = \frac{\mathrm{ln}(N)}{\lambda} + \mathcal{O}(N^0)
    \label{leadingtstar}
\end{equation}
 As well as fitting to the functional form~\eqref{func_form} we will also provide evidence for the scaling~\eqref{fastscrambler} by directly extracting the scrambling time from our numerical results for $S_m(t)$ using the definition~\eqref{scrambling_time}. We will demonstrate a strong numerical functional fit to the scaling $t_* \sim \ln(N)$.

We first present numerical evidence that the near-saturation behaviour of the averaged operator entanglement entropy is well described by \eqref{func_form}. It is convenient to define 
 \begin{equation}
 \Delta S_m(t) = \frac{S^{\mathrm{sat}}_m - S_m(t)}{N}
 \label{deviation}
 \end{equation}
\begin{figure}[h!]
\centerline{\includegraphics[scale= 0.55]{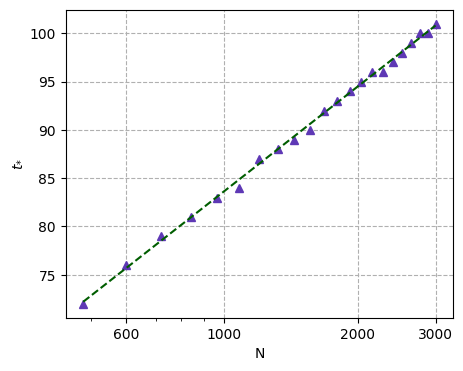}}
\caption{The scrambling time $t_*$ of a parallel processing super-Clifford circuit for different system sizes $N$.  The purple triangles correspond to the scrambling time extracted directly from numerical data for the averaged operator entanglement entropy $S_{1/4}(t)$ and~\eqref{scrambling_time} with $\epsilon = 10$. The green line is the line of best fit to the Ansatz $t_* = a \ln(N) + b$. }
\label{SaturationTime}
\end{figure}
 which describes the (normalised) deviation of the operator entanglement entropy from its saturation value. In Figure~\ref{StraightLine} we plot $\ln(\Delta S_{1/4})(t)$ for a range of $N \geq 1560$, with $S_{1/4}(t)$ averaged over $500$ iterations of the random circuit (recall our circuit is defined for $N$ divisible by $40$).  For the functional form $\eqref{func_form}$ then $\ln(\Delta S_{1/4}(t))$ is given by a straight line with an $N$ independent slope and intercept. The results in Figure~\ref{StraightLine} are clearly remarkably well described by such behaviour. We can quantify this by considering a straight-line fit to $\ln(\Delta S_{1/4})(t)$ for $N = 3000$, which is the black line in~Figure~\ref{StraightLine} . We find that this straight-line is a remarkably good fit, not only for the data corresponding to $N=3000$ but also for all values of $N$ in~Figure~\ref{StraightLine}, with $R^2$ goodness of fit in excess of $0.998$ for all such $N$. As explained above, the fact that the entropy is well described by the functional form~\eqref{func_form} implies that it exhibits fast scrambling.   %

Whilst the data illustrated above was computed for the case $m=1/4$ the functional form~\eqref{func_form} also well describes other subregions (i.e. choices of $m$) for large $N$. We find that the parameter $\alpha$ appearing in the functional form is a non-trivial function of $m$. Interestingly however the parameter $\lambda$ is approximately independent of $m$ - for instance, repeating the analysis above with $m$ taking on values  $1/3$, $1/4$ and $1/5$ we find that $\lambda$ takes values $0.0636$, $0.0614$, and $0.0630$. As a result the leading behaviour of the scrambling time at large-$N$ given by~\eqref{leadingtstar} appears independent of the subregion used to compute the operator entanglement.

The scaling form~\eqref{func_form} is highly instructive in understanding the near-saturation behaviour of the averaged operator entanglement entropy for large $N$. However it is interesting to also perform a direct extraction of the scrambling time from our numerical results. That is, after averaging the operator entanglement entropy $S_m(t)$ over $500$ iterations of the circuit  we can extract the scrambling time directly from~\eqref{scrambling_time}. Doing so requires a choice of the arbitrary cutoff $\epsilon$ - for the results presented here we take $\epsilon = 10$ (although we see broadly similar behaviour for other choices of $\epsilon$). In~Figure \ref{SaturationTime} we plot $t_{*}$ extracted using this method for a range of system sizes between $500$ and $3000$ qubits. These data points are then fitted using a two parameter non-linear fit with a logarithmic Ansatz:
\begin{equation}
    t_*(N) = a \mathrm{ln}(N) + b
\end{equation}
We find this Ansatz provides a good fit for the data with optimal parameters when $a=15.61$, and $b=-24.18$. The corresponding $R^2$-goodness of fit is $0.9989$, which providing strong evidence the circuit introduced in Section~\ref{sec:fastscrambler} is a fast scrambler.

To conclude this section, it is instructive to clarify the origin of the fast scrambling behaviour $t_* \sim \ln(N)$ in the operator entanglement of these circuits. It was argued in \cite{Bentsen_2019}, in the context of state entanglement, that a fast scrambling bound on state entanglement arises due to a slow-down in the rate of entropy production at late times. Specificially, it was argued that the entropy of macroscopic subregions is governed by (at best) a slow exponential decay towards maximal entanglement. The numerical results in this paper, specifically the fit to the functional form~\eqref{func_form} can be viewed as an explicit realisation of this behaviour in the context of the (averaged) operator entanglement of this parallel-processing super-Clifford circuits. 

\section{Out-of-time-Ordered Correlators}
\label{sec:OTOC}

In addition to operator entanglement, another important diagnostic of the scrambling of operators is provided by the out-of-time ordered correlator (OTOC) \cite{Hosur_2016, Roberts_2015}. In particular, the scrambling of an operator $W(t)$ can be seen by computing the quantity
 \begin{eqnarray}
 F(t) &\equiv& \frac{1}{2^N} \mathrm{Tr}(W(t)^{\dagger} V(0)^{\dagger} W(t)V(0))
 \label{OTOC}
 \end{eqnarray}
with $W(t) = U(t)^{\dagger}W(0)U(t)$ the Heisenberg time-evolution of $W(0)$, and $V(0)$ a distinct chosen operator. In the majority of studies in the literature, OTOCs are studied for traceless few-body operators $W(0), V(0)$. In sufficiently chaotic systems with large total Hilbert space dimension such OTOCs then decay to zero at late-times, which is the hallmark of scrambling \cite{Shenker_2014, Roberts_2015, Roberts:2016hpo, Stanford_2022, Yoshida:2017non}.

In this Section we will demonstrate that in super-Clifford circuits we are able to compute certain OTOCs that probe the scrambling of operators $W(t) \in \mathcal{S}$ using the super-stabiliser formalism reviewed in Section II. Furthermore, such OTOCs can be computed in polynomial time, and hence in super-Clifford circuits can be numerically studied in systems of thousands of qubits. We illustrate this by numerically computing OTOCs for the `parallel processing' super-Clifford circuit introduced in Section~\ref{sec:fastscrambler}. In all cases we find that the late time behaviour of the OTOC approaches a value indicating that $W(t)$ is scrambled in $\mathcal{S}$. 

\subsection{OTOCs in the super-Clifford formalism} 

We begin by demonstrating that certain OTOCs of the form~\eqref{OTOC} can be computed in polynomial time in super-Clifford circuits using the super-stabiliser formalism. In particular, we consider OTOCs where 
\begin{enumerate}
\item $W(t)$ is an operator in ${\cal S}$ generated by evolving a computational basis operator in $\mathcal{S}$ (e.g. $X_1 \dots X_N$) with a super-Clifford circuit. 
\item $V(0)$ is any super-Clifford operator (or arbitrary product of them). 
\end{enumerate}
%
%
In our explicit numerical simulations in Section~\ref{sec:numerics} we will choose $V(0)$ to be a few-body operator, i.e. acting on only a small number of qubits. However we note that one could also use the super-stabiliser formalism to efficiently compute OTOCs with $V(0)$ a many-body super-Clifford operator with support on ${\cal O} (N)$ qubits. 

The above assumptions have two crucial implications. Firstly, we have that these conditions the operators $W(t)$ and $V^{\dagger} W(t) V$ are both operators in $\mathcal{S}$. The OTOC can therefore be expressed as
\begin{eqnarray}
 F(t) &\equiv& \frac{1}{2^N} \mathrm{Tr}(W(t)^{\dagger} V(0)^{\dagger} W(t)V(0)) = \langle {\bf W} (t)| {\bf V^{\dagger}} {\bf W}(t) {\bf V} \rangle  
\label{innprod}
\end{eqnarray}
where 
\begin{equation}
\langle {\bf Q_1} | {\bf Q_2} \rangle = \frac{1}{2^{N}} \mathrm{Tr}(Q_1^{\dagger} Q_2). \nonumber 
\end{equation}
is the conventional inner-product between operators in $Q_1, Q_2 \in \mathcal{S}$. Secondly, with the above assumptions we have not only are $W(t)$ and $V^{\dagger} W(t) V$ both operators in $\mathcal{S}$, but further that they are stabiliser operators. The OTOC \eqref{innprod} therefore reduces to computing the inner product of two stabiliser states $\langle {\bf W} (t)| {\bf V^{\dagger}} {\bf W}(t) {\bf V} \rangle$. This can be achieved using the techniques of \cite{Aaronson_2004}, which introduced a standard algorithm for computing such inner products in terms of the stabiliser formalism. 


In particular, given two stabiliser operators $\Ket{\mathbf{Q_1}}$, $ \Ket{\mathbf{Q_2}}$ the inner product will be zero if the (super)-stabiliser groups $G_1, G_2$ share a generator that differs only by an overall sign. Otherwise, the inner product will be given by $2^{-k/2}$, where $k$ is the minimum over all sets of generators of $G_1, G_2$ of the number of distinct generators  \cite{Aaronson_2004}. In our context, taking $W(0) = X_1 \dots X_N$, the inner product is computed by re-expressing the OTOC as 
\begin{equation}
F(t)  = \langle {\bf 00...0} | {\bf \tilde{Q}_2}(t) \rangle \nonumber 
\label{otocapp2}
\end{equation}
with $ \lbrack {\bf \tilde{Q}_2} (t) \rangle$ the state corresponding to the operator $\tilde{Q}_2(t)=  U(t) V^{\dagger} U(t)^{\dagger} X_1 \dots X_N U(t) V U(t)^{\dagger}$. One then compares the (super)-stabiliser group of $\lbrack {\bf \tilde{Q}_2} (t) \rangle$ to that of $\Ket{00 \dots 0}$ as described above. In practice, this is achieved by combining the super-stabilisers of $\lbrack {\bf \tilde{Q}_2} (t) \rangle$ into a suitable super-stabiliser matrix and then performing Gaussian elimination to determine the minimum number of super-stabilisers that are distinct from those of $\Ket{00 \dots 0}$ \cite{Aaronson_2004}. 

%
%

The details of performing this Gaussian elimination and generalisation to other choices of $W(0)$ are discussed in Appendix~\ref{app:OTOCstab}. For the purposes of the main text the essential point is that, given the conditions 1 and 2 above, the OTOC computation amounts to determining the super-stabilisers of the operator $\lbrack {\bf \tilde{Q}_2} (t) \rangle$ for a given super-Clifford circuit $U(t)$, before performing Gaussian elimination on the resulting tableaux. As a result for super-Clifford circuits we have that such OTOCs can be computed in polynomial (in $N$) time.

\subsection{Scrambling in a subspace of operators}
\label{sec:latetime}

Before presenting numerical results for OTOCs in the super-Clifford circuit introduced in Section~\ref{sec:fastscrambler}, we wish to discuss the expected late-time behaviour of \eqref{OTOC} for operators $W(t) \in \mathcal{S}$. In particular, as we have emphasised, for super-Clifford circuits the scrambling of operators $W(t)$ takes places within a subspace $\mathcal{S}$ of the full Hilbert space of operators. We therefore wish to ask what value of the OTOC~\eqref{OTOC} indicates that an operator $W(t)$ is scrambled within this subspace of operators. 

We first recall standard results for understanding the late time behaviour of the OTOC, for the case where the scrambling of $W(t)$ is unconstrained - i.e. takes place in the full Hilbert space of operators. In this case, the late time behaviour is expected to be given by replacing $W(t)$ in~\eqref{OTOC} by a Haar random operator $W(U) = U^{\dagger} W U$. After averaging over $U$ one finds, at large total Hilbert space dimension, the result for traceless $W(0)$ with normalization $\mathrm{Tr}(W^2) = 2^{N}$ reduces to \cite{Roberts:2016hpo, Stanford_2022, Yoshida:2017non} 
\begin{equation}
F(t) =  |\langle V(0) \rangle|^2
\label{OTOClate1}
\end{equation}
where $\langle V(0) \rangle = \mathrm{Tr}(V(0))/2^{N}$. The OTOC~\eqref{OTOC} approaching the value~\eqref{OTOClate1} at late times is then the signature in the OTOC of the scrambling of the operator $W(t)$. 

For the case of super-Clifford circuits, where the scrambling of $W(t)$ is constrained to the subspace $\mathcal{S}$, it is no longer appropriate to average over Haar random unitaries since $W(U)$ generically will lie outside $\mathcal{S}$. However we note that an equivalent way of obtaining~\eqref{OTOClate1} is to replace $W(t)$ in \eqref{OTOC} by a maximally entangled operator between $A$, $\bar{A}$, with $A$ the domain on which $V(0)$ has non-trivial support~\cite{Dowling:2023hqc}. We can therefore obtain a generalisation of \eqref{OTOClate1} for our case of scrambling within the operator subspace $\mathcal{S}$ by now replacing $W(t)$ in \eqref{OTOC} by an operator $W_{\mathrm{max}}$ that is maximally entangled between $A, \bar{A}$, given the constraint it lies in $\mathcal{S}$. For such a maximally entangled operator the OTOC reduces to\footnote{This follows from the Schmidt decomposition $W_{\mathrm{max}} = \lambda \sum_{i=1}^{d_A} P_i \otimes W_i^{\bar{A}}$ with $\lambda^2 = 2^{N}/d_A$, $W_i^{\bar{A}}$ orthogonal operators on $\bar{A}$ with $\mathrm{Tr}(W_i^{\bar{A}}W_j^{\bar{A}}) = \delta_{ij}$, and we have assumed $d_A < d_{\bar{A}}$.}
\begin{equation}
F(t) = \sum_{i = 1}^{d_A} \frac{1}{d_A} \mathrm{Tr}(P_i V(0)^{\dagger} P_i V(0)). 
\label{otocaverage}
\end{equation}
where $P_i$ is proportional to a Pauli string on $A$ with $X$ or $Y$ on each site, normalised such that $\mathrm{Tr}(P_i P_j) = \delta_{i j}$. Note that such operators form a basis for the tensor factor of $\mathcal{S}$ that corresponds to region $A$. For example, if $A$ consists of $3$ qubits we have 
\begin{eqnarray}
P_1 &=& \frac{1}{\sqrt{8}} X_1 X_2 X_3,  \hspace{1.0cm} P_2 = \frac{1}{\sqrt{8}}  X_1 X_2 Y_3,  \hspace{1.0cm} P_3 = \frac{1}{\sqrt{8}}  X_1 Y_2 X_3, \hspace{1.0cm} P_4 = \frac{1}{\sqrt{8}}  X_1 Y_2 Y_3, \nonumber \\
P_5 &=& \frac{1}{\sqrt{8}} Y_1 X_2 X_3, \hspace{1.1cm} P_6 = \frac{1}{\sqrt{8}}  Y_1 X_2 Y_3, \hspace{1.1cm} P_7 = \frac{1}{\sqrt{8}} Y_1 Y_2 X_3, \hspace{1.1cm} P_8 =  \frac{1}{\sqrt{8}} Y_1 Y_2 Y_3, \nonumber 
\end{eqnarray}
The result~\eqref{otocaverage} can be computed directly for a given operator $V(0)$ and gives the expected late time behaviour of the OTOC for an operator $W(t)$ that is scrambled in $\mathcal{S}$, providing a generalisation of~\eqref{OTOClate1} to this setting.

\subsection{Numerical results for OTOCs} 
\label{sec:numerics}

\begin{figure*}[htp]
    \centering
    \includegraphics[width=.50\textwidth]{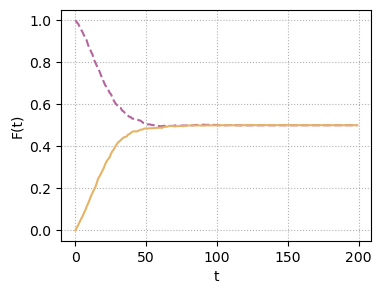}\hfill
     \includegraphics[width=.50 \textwidth]{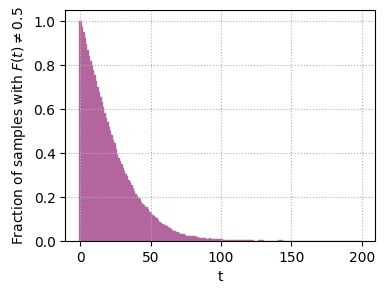}\hfill
    \caption{The left hand panel shows OTOCs~\eqref{OTOC} for $V(0) = C_3$ in a parallel processing super-Clifford circuit with $N=120$ qubits, averaged over $1000$ realisations of the circuit. The plots are shown for $W(0) = D_1 X_2 \dots X_N$ with $D_1 = X_1$ (mauve dashed line) and $D_1 =Y_1$ (yellow solid line). The right hand panel shows the fraction of realisations of the circuit for which $F(t)$ is not given by the scrambled value $F(t) = 1/2$ (out of $2000$ realisations with $D_1 = X_1$).}
    \label{OTOC_figures}
    \end{figure*}

We now wish to use the OTOC~\eqref{OTOC} to probe the scrambling of operators $W(t) \in \mathcal{S}$ in the `parallel processing' super-Clifford circuit introduced in Section~\ref{sec:fastscrambler}. We provide numerical results for several distinct choices of the operators $W(0), V(0)$ in the OTOC~\eqref{OTOC}, where $V(0)$ is a few-body super-Clifford gate acting on the first three qubits. In all cases our results are consistent with scrambling of $W(t)$ in $\mathcal{S}$, with the late time value of the OTOC approaching~\eqref{otocaverage}.

The numerical results presented in this Section were produced by using the Clifford simulator stim package~\cite{Gidney_2021} to track the time evolution of stabilisers of $\tilde{Q}_2(t)$, and then performing the Gaussian elimination discussed in Appendix~\ref{app:OTOCstab}.  We note that in an individual realisation of the circuit the OTOC, given by inner product of stabiliser states~\eqref{innprod}, takes discrete values of the form $0$ or $2^{-k/2}$, where $k$ is a positive integer. There are also large circuit-to-circuit fluctuations for the OTOC at early times. Hence to obtain smooth curves for the OTOC~\eqref{OTOC} we average over many iterations of the circuit.

\begin{figure*}[htp]
    \centering
       \includegraphics[width=.50\textwidth]{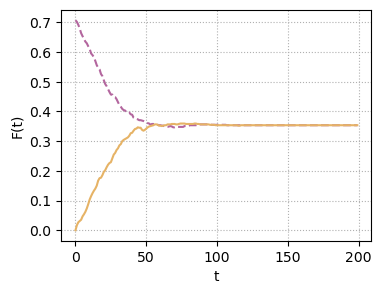}\hfill
        \includegraphics[width=.50 \textwidth]{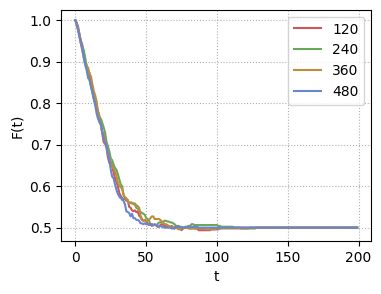}\hfill
    \caption{The left hand panel shows the OTOC averaged over $250$ realisations of a parallel processing super-Clifford circuit with $N=1000$ qubits, for operators $V(0) = C_3 T_3$ and $W(0) = D_1 \dots X_N $ with $D_1 = X_1$ (mauve dashed line) and $D_1 =Y_1$ (yellow solid line).  The right hand panel shows the OTOC, averaged over $250$ realisations, for $V(0) = C_3 $, $D_1 = X_1$ for $N = (120, 240, 360, 480)$ qubits.}
    \label{OTOC_figures2}
    \end{figure*}

Results for OTOCs for the circuit in Section~\ref{sec:fastscrambler} are presented in Figures~\ref{OTOC_figures} and~\ref{OTOC_figures2}. The left hand panel of Figure~\ref{OTOC_figures} displays the averaged OTOC in a system of $N=120$ qubits, where $V(0) = C_3$ (acting on the first three qubits). The two distinct lines in the plot correspond to different choices of initial computational basis operator, $W(0) = X_1 X_2 \dots X_N$ and $W(0) = Y_1 X_2 \dots X_N$. The plots demonstrate the expected features for scrambling of $W(t)$ in $\mathcal{S}$. In particular, the initial value of the OTOC depends on the initial computational basis operator $W(0)$. However at late times, for each choice of $W(t)$, the OTOC approaches $F(t) = 1/2$ which is the expected late time value~\eqref{otocaverage} indicating $W(t)$ is scrambled in $\mathcal{S}$. 

We emphasise that, whilst the plots in the left hand panel of Figure~\ref{OTOC_figures} are made by averaging over many realisations of the circuit, the fluctuations in the OTOC are very infrequent at late times, such that scrambling indeed takes place within a typical, individual realisation of a circuit. This is illustrated in the right hand plot of Figure~\ref{OTOC_figures}. Recall that in a given realisation of the circuit, the OTOC takes discrete values of $0$ or $2^{-k/2}$, with $k$ a positive integer. The plot in Figure~\ref{OTOC_figures} then shows the fraction of realisations of the circuit where $F(t)$ is not equal to the value $1/2$ (indicating scrambling) as a function of time. We find that at late times this fraction decreases towards zero, indicating scrambling in almost all realisations of the circuit, and that at $t=120$ the fraction of samples where $F(t) = 1/2$ is already given by 0.9985.

We have also computed the OTOC for different choices of the super-Clifford gate $V(0)$, for which we find similar results to the case $V(0) = C_3$. In particular, results for the averaged OTOC for a circuit with $N=1000$ qubits are shown in the left hand panel of~Figure~\ref{OTOC_figures2} for the case of $V(0) = T_3 C_3$. The two plots again correspond to the initial computational basis operators $W(0) = X_1 X_2 \dots X_N$ and $W(0) = Y_1 X_2 \dots X_N$. The same features of scrambling are present, with the OTOCs approaching the scrambled value $F(t) = 1/\sqrt{8}$ (from~\eqref{otocaverage}) at late times. 

An interesting aspect of our results is that, for large $N$, the averaged OTOCs we have computed are broadly independent of the number of qubits $N$ in our parallel-processing super-Clifford circuit. The is illustrated in the right hand panel of Figure~\ref{OTOC_figures2} which displays results for the averaged OTOC for $W(0) = X_1 \dots X_N$ and $V(0) = C_3 $ for a range of values of $N$ (similar behaviour is found for the other choices of $V(0)$ we have discussed). It is particularly noteworthy that this implies that the time taken for these averaged OTOCs to decay to approach their late time value~\eqref{otocaverage} is $N$ independent. This should be contrasted with the behaviour of OTOCs of few-body operators in fast-scrambling systems, which decay to~\eqref{OTOClate1} over the scrambling time $t_* \sim \ln(N)$. This difference can be traced to the fact that the OTOCs we can compute using the super-Clifford formalism involve a many-body operator $W(0)$ with support on all qubits. From our discussion in Section~\ref{sec:latetime}, the OTOC will reach its late time value when $W(t)$ becomes scrambled on the ${\cal O} (1)$ sites $A$ where $V$ has non-trivial support. Given that $W(0)$ has non-trivial support on $A$ and the parallel processing circuit entangles these sites with ${\cal O}(1)$ probability in a given timestep, then $W(t)$ should indeed become scrambled on these sites in an ${\cal O}(1)$ time.

\section{Discussion} We have studied the scrambling of operators in super-Clifford circuits \cite{Blake_2020}, for which the scrambling of a subset of operators $W(t)$ can be efficiently simulated on a classical computer.  Our results go beyond the initial studies of \cite{Blake_2020} in two keys ways. Firstly, we numerically investigated the speed at which operator entanglement is generated in a parallel-processing super-Clifford circuit. We found strong evidence that this system is a `fast scrambler', in the sense that the scrambling time (extracted from operator entanglement) scales as $t_* \sim \mathrm{ln}(N)$.  Secondly, we demonstrated that in super-Clifford circuits certain OTOCs involving $W(t)$ could be computed in polynomial time using the stabiliser formalism. We numerically studied examples of OTOCs in our parallel-processing super-Clifford circuit, and found that they exhibited the expected features of scrambling of $W(t)$. This represents a significant expansion in the types of probes of scrambling that can be efficiently computed in super-Clifford circuits, beyond the case of operator entanglement identified in \cite{Blake_2020}.

Throughout this paper we have been concerned with the scrambling of `many-body' operators $W(t)$ which have global support on our system of $N$ qubits. Whilst the majority of studies in the literature focus on scrambling of local few-body operators, the scrambling of similar global operators has previously been studied in~\cite{PhysRevA.94.040302}. In particular,~\cite{PhysRevA.94.040302} presented results for studies of OTOCs where both $V(0)$ and $W(0)$ can be viewed as a global product of rotation operators in an ensemble of spins.\footnote{Other studies of operators with global support include~\cite{Lin_2018, Zhou_2023, Kukuljan_2017}. The global operators studied in \cite{Zhou_2023} and \cite{Kukuljan_2017} are qualitatively different to those studied in this paper, as they can be expressed as sums of local operators.} In this paper we studied a slightly different form of the OTOC, where $W(0)$ was a global operator and $V(0)$ a local operator. Surprisingly we found that such OTOCs between many-body and local operators scrambled rapidly within an $\mathcal{O}(1)$ time in our parallel-processing super-Clifford circuit. It would be interesting to understand if this behaviour is generic for such OTOCs in fast scrambling systems. A further interesting direction is to use super-Clifford circuits to study more general types of OTOCs involving many-body operators. The method we have introduced to compute OTOCs in super-Clifford circuits can also be applied to OTOCs where both $W(0)$ and $V(0)$ are operators with global support, as in \cite{PhysRevA.94.040302}, and studying this is an interesting avenue for future work. 

An important motivation for future work is to understand if the results of \cite{Blake_2020} and this paper can lead to insights into other fast scrambling systems with a `large-$N$' limit, such as holographic quantum field theories or the SYK model.  As we have emphasised, the operators whose scrambling we can currently study using the super-Clifford formalism are `many-body operators'  with non-trivial support on each qubits - specifically linear combinations of Pauli strings with $X$ or $Y$ at each site. The majority of studies of scrambling in SYK models and holography so far (e.g. \cite{Shenker_2014, Roberts_2015,  RobertsStanford2018, Maldacena_2016, Kitaev_2018}) have taken place for a qualitatively distinct class of operators - few-body operators with non-trivial support only on a small number of qubits (and identities elsewhere).

A natural open question then remains whether the super-Clifford formalism can be adapted to study few-body operators - for instance whether one can identify gates that act as super-Clifford circuits in the subspace of operators formed by taking combinations of strings of $I$, $X$. General results on the super-Clifford formalism, however, imply that there cannot exist examples of such circuits that generate operator entanglement starting from a product of $I$s. This follows from the fact that the operator entanglement of $W(t)$ for any choice of computational basis operator $W(0)$ is identical under time evolution by a super-Clifford circuit, and that $ I_1 I_2 \dots I_N$ can never be entangled by unitary dynamics. We provide an alternative perspective on the result in Appendix~\ref{app:local}. Whilst these simplest attempts are not immediately successful, it would be extremely worthwhile to see if combining our approach with other techniques for simulating quantum dynamics classically can lead to circuits capable of simulating the scrambling of few-body operators. 

A question of more immediate interest may be to instead directly ask about the scrambling of many-body operators (as considered here and in \cite{Blake_2020}) in SYK models or holography.  Roughly speaking, in the context of SYK models featuring $N$ Majorana fermions, this would involve studying the scrambling of `baryonic' operators consisting of a product of $N$ fermions $\Psi_1 \dots \Psi_N$. Similarly it would be interesting to understand the holographic description of such operators - a starting point for which could be provided by recent work on `huge' operators in AdS/CFT whose conformal scaling dimensional is proportional to the central charge \cite{Abajian:2023jye, Abajian:2023bqv}.

\begin{acknowledgements}  AT acknowledges support from UK Engineering and Physical Sciences Research Council  (EP/SO23607/1). MB acknowledges support from UK Research and Innovation (UKRI) under the UK government's Horizon Europe guarantee (EP/Y00468X/1).  NL gratefully acknowledges support from the UK Engineering and Physical Sciences Research Council through grants EP/R043957/1, EP/S005021/1, EP/T001062/1. The data and code that supports the findings of this article are openly available at \cite{blake2024}. For the purpose of open access, the authors have applied a Creative Commons Attribution (CC BY) licence to any Author Accepted Manuscript version arising from this submission.
\end{acknowledgements}

\bibliography{bibliography}

\appendix

\section{Averaging scrambling time over individual realisation of a circuit}
\label{app:scramblingtime}

The results in Section \ref{sec:entanglement}, including the logarithmic growth of the scrambling time, were extracted from the operator entanglement entropy averaged over many realisations of our parallel-processing super-Clifford circuit. An alternative way of defining a typical scrambling time associated with such circuits is to first extract the scrambling time $t_*$ using \eqref{scrambling_time} for an individual realisation of the circuit, and then computing the average $\langle t_* \rangle$ over different realisations of the circuit. 

\begin{figure*}[htp]
\centering
\includegraphics[width=.5\textwidth]{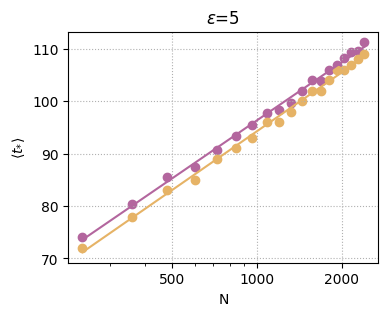}\hfill
\includegraphics[width=.5\textwidth]{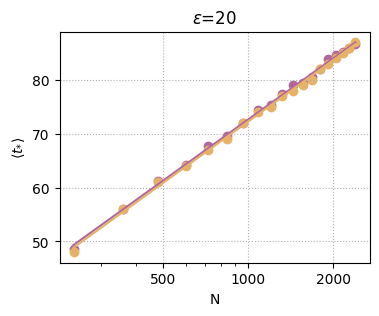}\hfill
\caption{The scrambling time $t_{*}$ of a parallel processing super-Clifford circuit for different numbers of qubits $N$, computed for alternative methods of averaging over realisations of the circuit. The yellow points are for the scrambling time extracted from the circuit averaged operator entanglement entropy $S_{1/4}(t)$ (over $60$ realisation of the circuit). The mauve points are obtained by first extracting the scrambling time for each realisation, and then averaging the scrambling times $\langle t_* \rangle$.  The yellow (mauve) line is the line of best fit through the yellow (mauve) data points with the Ansatz $t_{*} = a \mathrm{ln}(N) + b$.  The two plots correspond to different values of the cut-off $\epsilon$ used in ~\eqref{scrambling_time} to define the scrambling time. }
\label{average_scrambling_time}
\end{figure*}

In Figure \ref{average_scrambling_time} we compare these two methods of defining a typical scrambling time, and demonstrate that averaging $t_*$ over individual realisations of the circuit gives a similar behaviour for the scrambling time to the results in the main text extracted from the averaged entropy. In particular, the scaling of the scrambling time as $t_{*} \sim \mathrm{ln}(N)$ is clearly visible for both definitions of the typical scrambling time.

\section{Computation of OTOCs using stabiliser formalism}
\label{app:OTOCstab} 

The OTOC $F(t)$ in \eqref{OTOC} can be expressed as an inner product of super-stabiliser states
\begin{equation}
F(t)  = \langle \mathbf{Q_1} | \mathbf{Q_2} \rangle 
\label{otocapp}
\end{equation}
where $\Ket{Q_1}$, $\Ket{Q_2}$ are the state representation of the stabiliser operators $Q_1(t) = W(t)$, $Q_2(t) = V^{\dagger} W(t) V$. We now explain operationally how to compute this inner product of stabiliser states using the algorithm of \cite{Aaronson_2004}. 

First we consider the case where $W(0) = X_1 \dots X_N$. In this case we note that $F(t)$ can equivalently be expressed as the inner product 
\begin{equation}
F(t)  = \langle \mathbf{00...0} | \mathbf{\tilde{Q}_2} \rangle 
\label{otocapp2}
\end{equation}
with $\Ket{\tilde{Q}_2}$ the state corresponding to the operator $\tilde{Q}_2(t)=  U(t) V^{\dagger} U(t)^{\dagger} X_1 \dots X_N U(t) V U(t)^{\dagger}$. The inner product~\eqref{otocapp2} can then be computed from knowledge of the stabilisers of $\tilde{Q}_2(t)$ as follows. First, we represent the data of the stabilisers of $\tilde{Q}_2(t)$ by forming the $N \times 2 N$ matrix given by listing the $N$ binary vectors
\begin{equation}
    \tilde{\mathbf{v}} = (v_{1x}, \dots v_{Nx},  v_{1z}, \dots, v_{N z}). 
   \label{vectorapp}
    \end{equation}
We highlight that the ordering of the data in \eqref{vectorapp} is distinct from that used in the main text \eqref{vector} for computing the operator entanglement entropy.  Schematically the resulting $N \times 2N$ matrix ${\mathcal M}$ representing the super-stabilisers takes the form%
\begin{equation}
{\mathcal M} = \bigg( {\mathcal X} \bigg|  {\mathcal Z} \bigg) 
\label{matrixapp}
\end{equation}
with ${\mathcal X}$, ${\mathcal Z}$ the $N \times N$ matrices obtained by restricting the vectors \eqref{vectorapp} to their $\mathbf{X_i}$, $\mathbf{Z_i}$ components respectively. The  vertical bar in \eqref{matrixapp} is simply meant to highlight the separation of the matrix into two parts and should not be interpreted as implying an inner product.

The inner product \eqref{otocapp2} is then computed as follows. Firstly, if the binary matrix ${\mathcal X}$ is full rank (in arithmetic mod $2$), then the OTOC  $F(t) = 2^{-N/2}$. Secondly, if the matrix ${\mathcal X}$ has rank $k < N$ then Gaussian elimination (in arithmetic mod $2$) is performed on the matrix ${\mathcal M}$ to express it in row echelon form - i.e. until the final $N-k$ rows of $\mathcal{X}$ are identically zero. The row operations carried out in Gaussian elimination correspond to swapping or multiplying elements of the set of super-stabilisers, and the rows of the updated matrix correspond to a different generating set for the super-stabiliser group of the operator $\Ket{\tilde{Q}_2}$. It is also necessary to keep track of how the signs $s_{\alpha}$ of the super-stabilisers are affected by the row operations - in particular this can be non-trivial when multiplying two stabilisers of the form~\eqref{stab_form}. Once the Gaussian elimination has been performed the OTOC $F(t)$ is extracted as follows. Firstly. if any of the signs $s_{\alpha}$ of the final $N-k$ super-stabilisers (those with support only on ${\mathcal Z}$) are $1$ then the OTOC $F(t) = 0$. Else the OTOC is given by $F(t) = 2^{-k/2}$. 

It is straightforward to generalise the above discussion to a different choice of initial computational basis operator $W(0)$. Any such operator can be written as $W(0) = \pm \tilde{T}^{\dagger} X_1 \dots X_N \tilde{T}$ where $\tilde{T}$ is a product of super-Clifford $T$ gates. The inner product~\eqref{otocapp} is now equivalent to 
\begin{equation}
F(t)  = \langle \mathbf{00...0} | \mathbf{\tilde{Q}_3} \rangle 
\label{otocapp3}
\end{equation}
with $\Ket{\tilde{Q}_3}$ the state representation of the operator 
\begin{equation}
\tilde{Q}_3(t)=  \tilde{T} U(t) V^{\dagger} U(t)^{\dagger} \tilde{T}^{\dagger} X_1 \dots X_N \tilde{T} U(t) V U(t)^{\dagger} \tilde{T}^{\dagger} 
\end{equation}
The OTOC~\eqref{otocapp3} is then extracted by performing Gaussian elimination of the stabilisers of $\Ket{\tilde{Q}_3}$ in the manner described above. 

\section{Scrambling of few-body operators}\label{app:local}

As discussed in the main text, the simplest attempt to generalise the super-Clifford formalism to include few-body operators would be to identify gates $C$ which are closed and act on operators as super-Clifford gates within the subspace of operators given by the span of Pauli strings involving $I$ or  $X$ on each site. We argued in the main text that such circuits are not capable of generating operator entanglement starting from a single string of $I$s and $Xs$. We now prove a slightly stronger result that leads to the same conclusion - namely that any gate $C$ that acts as a a super-Clifford in this subspace must map a string of $I$s and $X$s to another string - i.e. in the state language where we identify $I = \Ket{0}$ and $X = \Ket{1}$ such a gate maps computational basis states to other computational basis states.    

To prove this we consider the state $\Ket{00...0}$, for which a choice of generating set for the super-stabiliser group is given by $\{\mathbf{Z_1},..., \mathbf{Z_n}\}$. Now any unitary $C$ fixes the identity under conjugation. If we let $\mathbf{C}$ represent the action of the unitary on operators in our subspace of strings of $I$s and $X$s, then $\mathbf{C}$ must fix the super-stabiliser group of $\Ket{00...0}$ when it acts through conjugation. This means that its action on each $Z_{\alpha}$ must just give back another element of the super-stabiliser group, i.e. we must have 
\begin{equation}
    \mathbf{C} \mathbf{Z_\alpha} \mathbf{C}^\dagger = \prod_{\gamma \in \Gamma} \mathbf{Z}_\gamma 
    \label{condition1}
\end{equation}
where $\Gamma \subset \{1, ..., n\}$.

We now consider the action of $\mathbf{C}$ on products of $I$s and $X$s.  In state language these are product states where every qubit is either in state $\Ket{0}$ or $\Ket{1}$ - i.e. computational basis states, whose super-stabiliser group is generated by  $ \{ (-1)^{s_1} \mathbf{Z_1},..., (-1)^{s_n} \mathbf{Z_n}\}$ with $s_{i}$ given by $0$ if there is an $I$ at site $i$ and $1$ else. The basic point is these super-stabilisers differ only from those of $\Ket{00...0}$ by (possible) overall signs, and hence we can deduce from condition~\eqref{condition1} how $\mathbf{C}$ acts on them. We therefore have 

\begin{align}
    \mathbf{C}((-1)^{s_{\alpha}} \mathbf{Z_\alpha})\mathbf{C}^\dagger &=(-1)^{s_{\alpha}} \mathbf{C} \mathbf{Z_\alpha} \mathbf{C}^\dagger \nonumber \\
    &= (-1)^{s_{\alpha}} \prod_{\gamma \in \Gamma} \mathbf{Z_\gamma}  \nonumber 
\end{align}
As such any product of $\Ket{0}$ and $\Ket{1}$ gets sent to a stabilizer state whose stabilizer group is generated entirely by $N$ independent products of $\pm \mathbf{Z_i}$. Such a stabiliser group necessarily corresponds to that of a computational basis state - this can be seen by noting that in representation of stabilisers by the stabiliser tableaux in \eqref{matrixapp} one has $\mathcal{X}=0$ and $\mathcal{Z}$ a full rank matrix. In this case Gaussian elimination can then be performed on $\mathcal{M}$ until $\mathcal{Z} = I_N$. We recall that Gaussian elimination corresponds to a change in the choice of basis for the super-stabiliser group - the above argument therefore shows this is generated by $\pm \mathbf{Z_i}$ and corresponds to that of a computational basis state. 

The above argument naturally generalises to cases where we consider larger subspaces of operators involving the identity. For example, one can consider trying to find gates $U$ that are closed and act as super-Clifford gates in the $3^{N}$ dimensional space of operators formed by the span of strings of $I$, $X$, or $Y$. At each site we now have a $3$ dimensional vector space which we can model as a qutrit system,  letting $I = \Ket{0}$, $X = \Ket{1}$, $Y = \Ket{2}$. One can now consider unitary dynamics $U$ on the underlying qubit Hilbert space subject to the constraint that $U$ should preserve the subspace of operator space spanned by strings of $I$, $X$ and $Y$ when it acts on such operators by conjugation and act as a super-Clifford within this subspace (in the sense of a qutrit super-Clifford operator). However the above argument naturally extends to this setting with only minor modifications - once again such a super-Clifford gates would map computational basis states to computational basis states (in the sense of $\Ket{0}, \Ket{1}, \Ket{2}$) and cannot generate operator entanglement starting from a product of $I$s, $X$s and $Y$s. Likewise, the same conclusion applies if one attempts to find gates that act as qudit super-Clifford gates (with $d=4$) in the full $4^{N}$ space of operators spanned by Pauli strings. 

\end{document}